\documentclass[11pt]{article}

\usepackage{times}
\usepackage{url, color, epsfig, subfigure, amsmath, amsthm, amssymb}

\usepackage{times}
\usepackage{url,amsmath, amssymb,epsfig,subfigure}

\usepackage{algorithm}
\usepackage[noend]{algorithmic}

\newcounter{listcounter}

\renewcommand{\thelistcounter}{\roman{listcounter}}
\newcommand{\descr}{\begin{list}{(\thelistcounter)}
{\usecounter{listcounter}
\setlength{\rightmargin}{0mm}}}
\newtheorem{lemma}{Lemma}[section]
\newtheorem{theorem}{Theorem}

\newif\ifpfsymb

\newenvironment{Pf}{\par\addvspace{6pt}\addtocounter{proof}{1}
       \noindent{\bf Proof:}}{{\ifnum\value{proof}>0
                \hfill\hbox{\mbox{}\hfill$\Box$}
		\addtocounter{proof}{-1}\par\addvspace{10pt}\else\par\fi
        }}

\makeatletter
\newcounter{algo}
\def\thealgo{\@arabic\c@algo}
\def\fps@algo{tbp}
\def\ftype@algo{1}
\def\ext@algo{loa}
\def\fnum@algo{Algorithm \thealgo}
\def\algo{\@float{algo}}
\def\endalgo{\end@float}
\makeatother

\makeatletter
\def\remlab#1{\@bsphack\if@filesw {\let\thepage\relax
   \def\protect{\noexpand\noexpand\noexpand}%
\xdef\@gtempa{\write\@auxout{\string
	   \newlabel{rem:#1}{{\thelemma}{\thepage}}}}}\@gtempa
            \if@nobreak \ifvmode\nobreak\fi\fi\fi\@esphack}
\def\deflab#1{\write\@auxout{\string
	\newlabel{def:#1}{{\thelemma}{\thepage}}}}
\makeatother

{\catcode`\@=11
\gdef\setft#1#2#3{%
\def\@oddfoot{%
{\setbox0=\hbox{#1}%
\setbox1=\hbox{#3}%
\ifdim\wd0>\wd1%
\dimen0=\wd0%
\box0\hfil#2\hfil\hbox to\dimen0{\hfil\hfil\box1}%
\else \dimen0=\wd1%
\hbox to\dimen0{\box0\hfil}\hfil#2\hfil\box1\fi%
}}}}

{\catcode`\@=11
\gdef\sethd#1#2#3{%
\def\@oddhead{\vbox{\hbox to\hsize{{#1}\hfil{#2}\hfil{#3}}%
\vspace{0.06in}%
\hbox to \hsize{\hrulefill}\vspace*{-0.09in}}}
\def\@evenhead{\@oddhead}
	}
}

\def\mysecn#1{\setcounter{equation}{0}
\section*{#1}\mark{#1}}

\def\thebibliography#1{\mysecn{References}
\addcontentsline{toc}{section}{References}\list
{[\arabic{enumi}]}{\settowidth\labelwidth{[#1]}\leftmargin\labelwidth
 \advance\leftmargin\labelsep
 \usecounter{enumi}}
 \def\newblock{\hskip .11em plus .33em minus .07em}
 \sloppy\clubpenalty4000\widowpenalty4000
 \sfcode`\.=1000\relax
 \small}

\def\complaint#1{}
\def\withcomplaints{
\newcounter{mycomplaints}
\def\complaint##1{\refstepcounter{mycomplaints}%
\ifhmode%
\unskip%
{\dimen1=\baselineskip \divide\dimen1 by 2 %
\raise\dimen1\llap{\tiny -\themycomplaints-}}\fi%
\marginpar{\tiny [\themycomplaints]: ##1}}%
}

\newcounter{printertype}
\def\figprint#1{
        \ifcase \theprintertype

		\begin{center}
                 \input{#1}
		\end{center}
              \or
                 \centerline{\psfig{figure=#1.ps}}
              \else
                 \vspace*{1in}
        \fi}

\makeatletter
\long\def\@myfootnotetext#1{\insert\footins{\footnotesize
    \interlinepenalty\interfootnotelinepenalty 
    \splittopskip\footnotesep
    \splitmaxdepth \dp\strutbox \floatingpenalty \@MM
    \hsize\columnwidth \@parboxrestore
   \edef\@currentlabel{\csname p@footnote\endcsname\@thefnmark}\@makemyfntext
    {\rule{\z@}{\footnotesep}\ignorespaces
      #1\strut}}}

\def\myfootnotetext{\@ifnextchar
[{\@xfootnotenext}{\xdef\@thefnmark{\thempfn}\@myfootnotetext}}

\long\def\@makemyfntext#1{\parindent 5mm #1}

\newcounter{proof}
\def\@@meqncr{\let\@tempa\relax
    \ifcase\@eqcnt \def\@tempa{& & &}\or \def\@tempa{& &}
      \else \def\@tempa{&}\fi
     \@tempa $\Box$\addtocounter{proof}{-1}
     \global\@eqnswtrue\global\@eqcnt\z@\cr}
\@namedef{meqnarray*}{\def\@eqncr{\nonumber\@seqncr}\eqnarray}
\@namedef{endmeqnarray*}{\global\@eqcnt\tw@\@@meqncr\egroup
      \global\advance\c@equation\m@ne$$\global\@ignoretrue}

\@namedef{meqnarray}{\eqnarray}
\@namedef{endmeqnarray}{\@@meqncr\egroup
        \global\advance\c@equation\m@ne$$\global\@ignoretrue}
\def\mequation{$$\global\@ignoretrue}

\@namedef{Pf*}{\Pf}
\@namedef{endPf*}{\par\addvspace{10pt}}

\makeatother

\def\Ex{\mathop{\rm \textbf{E}}}

\def\P{\ensuremath{\mathcal{P}}}

\def\R{\ensuremath{\mathcal{R}}}

\newcommand{\myproof}{\textit{Proof. }}

\newcommand{\remove}[1]{}

\newsavebox{\smallProofsym}                            
\savebox{\smallProofsym}                               %
{
\begin{picture}(6,6)                                   %
\put(0,0){\framebox(6,6){}}                            %
\put(0,2){\framebox(4,4){}}                            %
\end{picture}                                          %
}   

\begin{document}
        
\title{A Note on the Size-Sensitive Packing Lemma}
\date{}

\author{Nabil H. Mustafa\footnote{Universit\'e Paris-Est, Laboratoire d'Informatique Gaspard-Monge, Equipe A3SI, ESIEE Paris. E-email: mustafan@esiee.fr. The work of Nabil H. Mustafa
in this paper has been supported by the grant ANR SAGA (JCJC-14-CE25-0016-01).}}

\maketitle
 
\begin{abstract}
We show that the size-sensitive packing lemma
follows from a simple modification of the standard proof, due to Haussler and simplified by Chazelle, of the packing lemma.
\end{abstract}
 
\section{Introduction}
\label{sec:intro}

In 1995 Haussler~\cite{H95} proved the following interesting theorem\footnote{In what follows below, we refer the reader to Matousek's textbook~\cite[Section 5.3]{M99} for notations and proofs.}.

\begin{theorem}[Packing lemma {\cite[Lemma 5.14, p. 156]{M99}}]
Let $(X, \P)$ be a set-system on $n$ elements, and with VC-dimension at most $d$.
Let $\delta$ be an integer, $1 \leq \delta \leq n$, 
such that  $|\Delta(R, S)|\geq \delta$  for every $R, S \in \P$,
where $\Delta(R, S) = (R \setminus S) \cup (S \setminus R)$.
Then $|\P| = O( (n/\delta)^d )$.
\label{thm:1}
\end{theorem}

Haussler's proof is a beautiful application of the probabilistic method,
and was simplified by Chazelle~\cite{C92}.
We refer to the discussion in~\cite{M99} for motivations and applications.
Recently much effort has been devoted to finding 
size-sensitive generalizations of this result.
After a series of partial
bounds~\cite{E14, MR14}, the following
statement has been recently established in~\cite{DEG15},
via \emph{two} different proofs (one building on Haussler's original proof
while the other extends Chazelle's proof):

\begin{theorem}[Size sensitive packing lemma]
Let $(X, \P)$ be a set-system on $n$ elements, 
and let $d, d_1, k, \delta > 0$ be  integers.
Assume the system has VC-dimension at most $d$.
Further, assume that for any set $Y \subseteq X$ the
number of sets in $\P_{|Y}$ of size at most $r$ is  at most
$f( |Y|, r ) = O( |Y|^{d_1} r^{d-d_1} )$.
If $|\Delta(R, S)|\geq \delta$  for every $R, S \in \P$
and $|S| \leq k$ for all $S \in \P$, then 
$|\P| = O( n^{d_1} k^{d-d_1} /\delta^d )$. 
\label{thm:2}
\end{theorem}
 
The objective of this present note is to point out
that Theorem~\ref{thm:2}, with a simple trick, is an immediate
consequence of the textbook proof~\cite{M99} of Theorem~\ref{thm:1}.

\paragraph{Haussler and Chazelle's Proof.} We rewrite the main step in their proof in a slightly more general form:

\begin{lemma}[{\cite[Proof of Packing Lemma 5.14, pp. 157--159]{M99}}]
Let $(X, \P)$ be a set-system on $n$ elements. 
Let $d, \delta$ be two integers such that
the VC-dimension of $\P$ is at most $d$,
and $|\Delta(S, T)| \geq \delta$ for all $S, T \in \P$.
Then
$$ |\P| \leq 2 \cdot \Ex[|\P_{|A'}|], \text{where $A'$ is a random
sample of size $\frac{4dn}{\delta}-1$}.$$
\label{lemma:1}
\end{lemma}
\myproof
Let $W, W_1, A, s = \frac{4dn}{\delta}$ 
be precisely as defined in the 
proof in the textbook~\cite[Proof of Packing Lemma 5.14, pp. 157--159]{M99},
where the following relations are proven:
$i)$ $2d|\P| \geq \Ex[W] = s \cdot \Ex[W_1]$, where the expectation
is over the choice of $A$, and 
$ii)$ $\Ex[W_1 | A' = Y] \geq \frac{\delta}{n} \big( |\P| - |\P_{|Y}| \big)$,
where  $A'$ is set to a fixed $Y$, and the expectation is over
the choice of a random element in $X \setminus Y$. Together they imply the lemma:
%
$$2d|\P| \geq \Ex[W] = s \cdot \Ex[W_1] \geq s \cdot \Big( \sum_{\substack{Y \subseteq X \\ |Y| = s-1}} \frac{\delta}{n} \big( |\P| - |\P_{|Y}| \big) \cdot \Pr[A'=Y] \Big) = 4d |\P| - 4d\Ex[|\P_{|A'}].$$ 
\qed

The proof in~\cite[Proof of Packing Lemma 5.14, pp. 157--159]{M99}
uses the primal shatter lemma to conclude the proof
of Theorem~\ref{thm:1}: $|\P_{|A'}| = O( (4dn/\delta)^d)$ for any $A'$ of size $s-1$.
Now we show that the proof of Theorem~\ref{thm:2} is also a similar step away,
by using instead the size-sensitive bound given in the assumption.

\paragraph{Proof of Theorem~\ref{thm:2}.} 

Let $A' \subseteq X$ be a random sample of size $\frac{4dn}{\delta}-1$.
Let $\P_1 = \{ S \in \P \text{ s.t. } |S \cap A'| \geq 3 \cdot 4dk/\delta \}$.
Note that $\Ex[|S \cap A'|] \leq 4dk/\delta$ as $|S| \leq k$ for all $S \in \P$.
By Markov's inequality, for any $S \in \P$, $\Pr[ S \in \P_1] = \Pr[ |S \cap A'| > 3 \cdot 4dk/\delta ] \leq 1/3$.
Thus
$$ \Ex[|\P_{|A'}|] \leq \Ex[|\P_1|] + \Ex[|(\P \setminus \P_1)_{|A'}|] \leq
\sum_{S \in \P} \Pr[ S \in \P_1 ] +   f(|A'|, 12dk/\delta) \leq \frac{|\P|}{3} + 
O\big( (\frac{4dn}{\delta})^{d_1} (\frac{12dk}{\delta})^{d-d_1} \big)$$
where the projection size of $\P \setminus \P_1$ to $A'$
is bounded by  $f(\cdot,\cdot)$. Applying
Lemma~\ref{lemma:1} finishes the proof.
\qed

\paragraph{Discussion.}

Besides a dramatically shorter proof,
%
our proof also improves
the constants in the bounds. Furthermore, we have shown that it follows
without any modification or addition to the Chazelle-Haussler proof. 
The
somewhat subtle  key idea
that was missed by   earlier work~\cite{DEG15, E14} is that it is fine if
$\Ex[|\P_{|A'}|]$ is bounded in terms of $c|\P|$ for a small-enough constant
$c$, as in any case
it would be absorbed by the LHS of the equation in Lemma~\ref{lemma:1}.
This allows us to replace the complicated technical machinery developed
in earlier work (iterative processes, Chernoff bounds for hypergeometric series, complicated probabilistic computations) for over 20 pages-long proofs
by a mere Markov's inequality.


\remove{
instead of using the shatter function bound 
to get Theorem~\ref{thm:1}, use
the assumed shatter function bound $f(\cdot, \cdot)$ to Theorem~\ref{thm:2}.

 (which presents Haussler's proof as simplified by Chazelle~\cite{C92}). 
The key statement that they proved is the following~(\cite{M99}, page 159):
\begin{equation}
\label{eq:1}
\end{equation}
From Equation~(\ref{eq:1}), the required bound follows trivially: by the
assumption (or the primal shatter lemma), $|\R_{|A}| = O( (4dn/\delta)^d)$
for any such $A$, and we're done.

\section{Conclusions}

In this note, 
}

\vspace{-0.2in}

\bibliographystyle{plain}
\bibliography{shallowpackinglemma}{}

\end{document}